\RequirePackage{lineno}
\documentclass[aps,prl,twocolumn,superscriptaddress,showpacs,preprintnumbers,amsmath,amssymb]{revtex4-1}

\usepackage{graphicx} 
\usepackage{dcolumn}  
\usepackage{epstopdf}
\usepackage{epsfig}
\usepackage{color}
\usepackage[dvipdfm, bookmarks]{hyperref}

\graphicspath{{ps}}

\begin{document}

\preprint{\vbox{ \hbox{   }
}}

\title{ \quad\\[1.0cm]First Determination of Level Structure of an $sd$-Shell Hypernucleus, $\rm {^{19}_{\Lambda}F}$ }

\noaffiliation
\affiliation{Department of Physics and Astronomy, Seoul National University, Seoul 08826, Korea}
\affiliation{Research Center for Nuclear Physics (RCNP), Osaka University, Ibaraki 567-0047, Japan}
\affiliation{Department of Physics, Korea University, Seoul 02841, Korea}
\affiliation{Department of Physics, Tohoku University, Sendai 980-8578, Japan}
\affiliation{Institute of Particle and Nuclear Studies (IPNS), High Energy Accelerator Research Organization (KEK), Tsukuba 305-0801, Japan}
\affiliation{Department of Physics, Kyoto University, Kyoto 606-8502, Japan}
\affiliation{Joint Institute for Nuclear Research, Dubna, Moscow Region 141980, Russia}
\affiliation{INFN, Sezione di Torino, via P. Giuria 1, 10125 Torino, Italy}
\affiliation{Advanced Science Research Center (ASRC), Japan Atomic Energy Agency (JAEA), Tokai, Ibaraki 319-1195, Japan}
\affiliation{Department of Physics, Osaka University, Toyonaka 560-0043, Japan}
\affiliation{Korea Research Institute of Standards and Science (KRISS), Daejeon 34113, Korea}
\affiliation{Faculty of Education, Gifu University, Gifu 501-1193, Japan}
\affiliation{Research Center of Nuclear Science and Technology (RCNST) and School of Physics and Nuclear Energy Engineering, Beihang University, Beijing 100191, China}

\author{S.~B.~Yang} \altaffiliation{Present Address: Research Center for Nuclear Physics (RCNP), Osaka University, Ibaraki 567-0047, Japan} \affiliation{Department of Physics and Astronomy, Seoul National University, Seoul 08826, Korea} \affiliation{Research Center for Nuclear Physics (RCNP), Osaka University, Ibaraki 567-0047, Japan} 
\author{J.~K.~Ahn}\affiliation{Department of Physics, Korea University, Seoul 02841, Korea} 
\author{Y.~Akazawa}\affiliation{Department of Physics, Tohoku University, Sendai 980-8578, Japan} 
\author{K.~Aoki}\affiliation{Institute of Particle and Nuclear Studies (IPNS), High Energy Accelerator Research Organization (KEK), Tsukuba 305-0801, Japan} 
\author{N.~Chiga}\affiliation{Department of Physics, Tohoku University, Sendai 980-8578, Japan} 
\author{H.~Ekawa}\affiliation{Department of Physics, Kyoto University, Kyoto 606-8502, Japan} 
\author{P.~Evtoukhovitch}\affiliation{Joint Institute for Nuclear Research, Dubna, Moscow Region 141980, Russia} 
\author{A.~Feliciello}\affiliation{INFN, Sezione di Torino, via P. Giuria 1, 10125 Torino, Italy} 
\author{M.~Fujita}\affiliation{Department of Physics, Tohoku University, Sendai 980-8578, Japan} 
\author{S.~Hasegawa}\affiliation{Advanced Science Research Center (ASRC), Japan Atomic Energy Agency (JAEA), Tokai, Ibaraki 319-1195, Japan} 
\author{S.~Hayakawa}\affiliation{Department of Physics, Osaka University, Toyonaka 560-0043, Japan} 
\author{T.~Hayakawa}\affiliation{Department of Physics, Osaka University, Toyonaka 560-0043, Japan} 
\author{R.~Honda}\affiliation{Department of Physics, Osaka University, Toyonaka 560-0043, Japan} 
\author{K.~Hosomi}\affiliation{Advanced Science Research Center (ASRC), Japan Atomic Energy Agency (JAEA), Tokai, Ibaraki 319-1195, Japan} 
\author{S.~H.~Hwang}\affiliation{Korea Research Institute of Standards and Science (KRISS), Daejeon 34113, Korea} 
\author{N.~Ichige}\affiliation{Department of Physics, Tohoku University, Sendai 980-8578, Japan} 
\author{Y.~Ichikawa}\affiliation{Advanced Science Research Center (ASRC), Japan Atomic Energy Agency (JAEA), Tokai, Ibaraki 319-1195, Japan} 
\author{M.~Ikeda}\affiliation{Department of Physics, Tohoku University, Sendai 980-8578, Japan} 
\author{K.~Imai}\affiliation{Advanced Science Research Center (ASRC), Japan Atomic Energy Agency (JAEA), Tokai, Ibaraki 319-1195, Japan} 
\author{S.~Ishimoto}\affiliation{Institute of Particle and Nuclear Studies (IPNS), High Energy Accelerator Research Organization (KEK), Tsukuba 305-0801, Japan} 
\author{S.~Kanatsuki}\affiliation{Department of Physics, Kyoto University, Kyoto 606-8502, Japan} 
\author{S.~H.~Kim}\affiliation{Department of Physics, Korea University, Seoul 02841, Korea} 
\author{S.~Kinbara}\affiliation{Faculty of Education, Gifu University, Gifu 501-1193, Japan} 
\author{K.~Kobayashi}\affiliation{Department of Physics, Osaka University, Toyonaka 560-0043, Japan} 
\author{T.~Koike}\affiliation{Department of Physics, Tohoku University, Sendai 980-8578, Japan} 
\author{J.~Y.~Lee}\affiliation{Department of Physics and Astronomy, Seoul National University, Seoul 08826, Korea} 
\author{K.~Miwa}\affiliation{Department of Physics, Tohoku University, Sendai 980-8578, Japan} 
\author{T.~J.~Moon}\affiliation{Department of Physics and Astronomy, Seoul National University, Seoul 08826, Korea} 
\author{T.~Nagae}\affiliation{Department of Physics, Kyoto University, Kyoto 606-8502, Japan} 
\author{Y.~Nakada}\affiliation{Department of Physics, Osaka University, Toyonaka 560-0043, Japan} 
\author{M.~Nakagawa}\affiliation{Department of Physics, Osaka University, Toyonaka 560-0043, Japan} 
\author{Y.~Ogura}\affiliation{Department of Physics, Tohoku University, Sendai 980-8578, Japan} 
\author{A.~Sakaguchi}\affiliation{Department of Physics, Osaka University, Toyonaka 560-0043, Japan} 
\author{H.~Sako}\affiliation{Advanced Science Research Center (ASRC), Japan Atomic Energy Agency (JAEA), Tokai, Ibaraki 319-1195, Japan} 
\author{Y.~Sasaki}\affiliation{Department of Physics, Tohoku University, Sendai 980-8578, Japan} 
\author{S.~Sato}\affiliation{Advanced Science Research Center (ASRC), Japan Atomic Energy Agency (JAEA), Tokai, Ibaraki 319-1195, Japan} 
\author{K.~Shirotori}\affiliation{Research Center for Nuclear Physics (RCNP), Osaka University, Ibaraki 567-0047, Japan} 
\author{H.~Sugimura}\affiliation{Advanced Science Research Center (ASRC), Japan Atomic Energy Agency (JAEA), Tokai, Ibaraki 319-1195, Japan} 
\author{S.~Suto}\affiliation{Department of Physics, Tohoku University, Sendai 980-8578, Japan} 
\author{S.~Suzuki}\affiliation{Institute of Particle and Nuclear Studies (IPNS), High Energy Accelerator Research Organization (KEK), Tsukuba 305-0801, Japan} 
\author{T.~Takahashi}\affiliation{Institute of Particle and Nuclear Studies (IPNS), High Energy Accelerator Research Organization (KEK), Tsukuba 305-0801, Japan} 
\author{H.~Tamura}\affiliation{Department of Physics, Tohoku University, Sendai 980-8578, Japan} 
\author{K.~Tanida}\affiliation{Advanced Science Research Center (ASRC), Japan Atomic Energy Agency (JAEA), Tokai, Ibaraki 319-1195, Japan} 
\author{Y.~Togawa}\affiliation{Department of Physics, Tohoku University, Sendai 980-8578, Japan} 
\author{Z.~Tsamalaidze}\affiliation{Joint Institute for Nuclear Research, Dubna, Moscow Region 141980, Russia} 
\author{M.~Ukai}\affiliation{Department of Physics, Tohoku University, Sendai 980-8578, Japan} 
\author{T.~F.~Wang}\affiliation{Research Center of Nuclear Science and Technology (RCNST) and School of Physics and Nuclear Energy Engineering, Beihang University, Beijing 100191, China} 
\author{T.~O.~Yamamoto}\affiliation{Department of Physics, Tohoku University, Sendai 980-8578, Japan} 

\collaboration{J-PARC E13 Collaboration}

\begin{abstract}
  We report on the first observation of $\gamma$ rays emitted from an $sd$-shell hypernucleus, $\rm ^{19}_{\Lambda}F$. The energy spacing between the ground state doublet, $1/2^{+}$ and $3/2^{+}$ states, of $\rm ^{19}_{\Lambda}F$ is determined to be $\rm 315.5 \pm 0.4 (stat) ^{+0.6} _{-0.5} (syst)~keV$ by measuring the $\gamma$-ray energy from the $M1(3/2^{+} \rightarrow 1/2^{+})$ transition. In addition, three $\gamma$-ray peaks were observed and assigned as $E2(5/2^{+} \rightarrow 1/2^{+})$, $E1(1/2^{-} \rightarrow 1/2^{+})$, and $E1(1/2^{-} \rightarrow 3/2^{+})$ transitions. The excitation energies of the $5/2^{+}$ and $1/2^{-}$ states are determined to be $\rm 895.2 \pm 0.3 (stat) \pm 0.5 (syst)~keV$ and $\rm 1265.6 \pm 1.2 (stat) ^{+0.7}_{-0.5} (syst)~keV$, respectively. It is found that the ground state doublet spacing is well described by theoretical models based on existing $s$- and $p$-shell hypernuclear data.
\end{abstract}


\maketitle

\tighten

{\renewcommand{\thefootnote}{\fnsymbol{footnote}}}
\setcounter{footnote}{0}


  Spectroscopy of $\Lambda$ hypernuclei has played essential roles in the recent trend of nuclear physics extending nuclear forces and nuclear systems into baryon-baryon interactions and baryonic many body systems~\cite{Gal_hypernucleus, Hiyama_hypernucleus, Hashimoto_hypernucleus}. Since the $\Lambda N$ scattering data is quite limited due to the short lifetime of $\Lambda$ hyperon, experimental data of structure of $\Lambda$ hypernuclei have also been used to extract information on $\Lambda N$ interaction. In particular, precise level data obtained via $\gamma$-ray spectroscopy for $s$- and $p$-shell hypernuclei have revealed strengths of the spin-dependent $\Lambda N$ interactions, by comparing them with those calculated from assumed baryon-baryon interactions~\cite{Tamura_hyperball_7_lambda_li, Akikawa_hyperball_9_lambda_be, Ukai_hyperball_16_lambda_o, Ukai_hyperball_7_lambda_li, Ukai_hyperball_16_lambda_o_15_lambda_n, Yamamoto_hyperball_4_lambda_he}. It is also found that the phenomenological spin-dependent interaction parameters determined from a few $p$-shell hypernuclear levels successfully reproduce almost all the $p$-shell hypernuclear level data~\cite{millener_theoretical_study}. \par
 It is interesting to ask whether the $\Lambda N$ spin-dependent interactions which successfully describe $s$- and $p$-shell hypernuclei can also be applied to heavier hypernuclei, since the distance and the overlap between a $\Lambda$ in the $0s$ orbit and valence nucleons in the outermost orbit are quite different in $s$-, $p$-, and $sd$-shell hypernuclei. Besides the two-body $\Lambda N$ interaction, the three-body $\Lambda N N$ interaction plays a particularly important role in the structure of hypernuclei~\cite{Akaishi_theoretical_study, Wirth_three_body_study} due to the small $\Lambda$-$\Sigma$ mass difference giving a large $\Lambda$-$\Sigma$ mixing in a hypernucleus. Thus, experimental data of various hypernuclei beyond $s$- and $p$-shell hypernuclei have been anticipated. Extension of spectroscopic study from $s$- and $p$-shell hypernuclei to $sd$-shell and heavier hypernuclei will allow us to test our theoretical frameworks for hypernuclear structure and our knowledge of the $\Lambda N$ interaction. \par

\begin{figure}[t]
\includegraphics[width=0.5\textwidth]{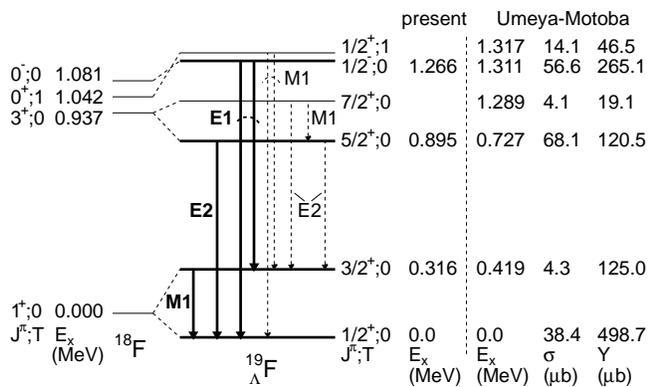}
\caption{Low-lying level scheme and $\gamma$ transitions of $\rm ^{19}_{\Lambda}F$. The ``present'' indicates the measured excitation energy and the thick arrows indicate observed transitions in this experiment. Dashed arrows show possible other transitions expected in this level scheme. Results of theoretical calculation by Umeya and Motoba~\cite{umeya_theoretical_study} are also shown. $\sigma$ is the production cross section by $(K^{-}, \pi^{-})$ reaction at 1.8 ${\rm GeV}/c$, integrated for $0^{\circ} - 12^{\circ}$ in the laboratory frame. $Y$ is the sum of the yield of the direct production and the population via $\gamma$-ray cascade from higher states below $E_{x}=6~\rm MeV$.}
\label{fig:scheme}
\end{figure}

 This letter reports the first experimental investigation of $sd$-shell hypernuclear structure via precise $\gamma$-ray spectroscopy using germanium (Ge) detectors. We selected $\rm ^{19}_{\Lambda} F$ as the first target of $sd$-shell hypernuclei, because $\rm {^{19}_{\Lambda}F}$ has a structure of ${\rm ^{16}O} + p + n + \Lambda$, being similar to the well known $\rm {^{7}_{\Lambda}Li}$ (${\rm ^{4}He} + p + n + \Lambda$)~\cite{Tamura_hyperball_7_lambda_li, Ukai_hyperball_7_lambda_li, Tanida_hyperball_7_lambda_li}. The ground state ($J^{\pi}=1^{+}$) of $\rm ^{18}F$ is split in $\rm ^{19}_{\Lambda}F$ hypernucleus by the spin of the additional $\Lambda$ as shown in Fig.~\ref{fig:scheme}. The energy spacing between the members of the ground state doublet, $1/2^{+}$ and $3/2^{+}$, is largely determined by the spin-spin $\Lambda N$ interaction because of the dominant $S=1$ and $L=0$ structure of $\rm ^{18}F$ $(1^{+})$, as in the ground state doublet of $\rm {^{7}_{\Lambda}Li}$ $(3/2^{+},1/2^{+})$. By measuring the $\gamma$-ray energy of the spin-flip $M1$ transition ($\rm ^{19}_{\Lambda}F(3/2^{+} \to 1/2^{+})$), the strength of the effective spin-spin interaction in the $sd$-shell hypernucleus can be obtained, and it can be directly compared with the effective spin-spin interaction for the $s$- and $p$-shell hypernuclei. The structure of $\rm ^{19}_{\Lambda} F$ has been theoretically studied via shell model~\cite{millener_theoretical_study, umeya_theoretical_study}; in Ref.~\cite{umeya_theoretical_study} the cross sections of $\rm ^{19}_{\Lambda} F$ excited states via the $(K^{-}, \pi^{-})$ reaction and their $\gamma$-transition strengths are also calculated.\par

 The experiment (J-PARC E13: $\gamma$-ray spectroscopy of $\rm {^{4}_{\Lambda}He}$ and $\rm {^{19}_{\Lambda}F}$) was performed at the K1.8 beam line in the J-PARC Hadron Experimental Facility~\cite{agari_j_parc_hadron_facility}. We have already reported the result of $\rm {^{4}_{\Lambda}He}$ $\gamma$-ray spectroscopy~\cite{Yamamoto_hyperball_4_lambda_he}. The experimental setup of the $\rm {^{19}_{\Lambda}F}$ study was the same as in the $\rm {^{4}_{\Lambda}He}$ case, except for the beam momentum and the target. \par

The $\rm {^{19}_{\Lambda}F}$ hypernuclei were produced by the ${\rm ^{19}F}(K^{-}, \pi^{-})$ reaction with a 1.8 ${\rm GeV}/c$ kaon beam and a 20 $\rm g/cm^{2}$-thick liquid $\rm {CF_{4}}$ target. The beam line spectrometer and the Superconducting Kaon Spectrometer (SKS) were used to identify $\rm {^{19}_{\Lambda}F}$ production~\cite{takahashi_spectrometer}, and at the same time $\gamma$ rays were detected with a Ge detector array, Hyperball-J~\cite{koike_hyperball_j}. The intensity of the kaon beam was typically $3.5 \times 10^{5}$ per one spill (2 s) occurring every 6 s and a typical $K^{-}/\pi^{-}$ ratio was 2.5. A total of $6.3 \times 10^{10}$ kaons was irradiated on the target. \par

The momenta and trajectories of beam kaons were measured by the beam line spectrometer. The beam kaons were identified at the trigger level by two aerogel \v{C}erenkov (AC) counters installed in front of the target. A misidentification probability of beam kaons was less than 1$\%$. A muon filter and a $\pi^{0}$ veto counter were used to reject decays of beam kaons, $K^{-} \rightarrow \mu^{-} {\bar{\nu}}_{\mu}$ and $K^{-} \rightarrow \pi^{-} \pi^{0}$, respectively. These veto counters made the trigger rate sufficiently low. For the outgoing pions, the SKS was used for measuring the momenta and trajectories. The outgoing pions were identified by an AC counter, installed just downstream the target, at the trigger level and by the time-of-flight in the off-line analysis. Through the off-line analysis, pions were well separated from other particles except for muons from the $K^{-}$ beam decay. More detailed descriptions of the experimental setup can be found in Ref.~\cite{Yamamoto_hyperball_4_lambda_he, yang_e13_proceeding}. \par

The reconstructed momenta of beam and outgoing particles were calibrated to reproduce the mass of $\Sigma^{+}$~\cite{PDG} and the $\Lambda$ binding energy of the $\rm ^{12}_{\Lambda}C$ ground state~\cite{Gogami_12LC} produced by the $(K^{-},\pi^{-})$ reaction with a $\rm CH_{2}$ target. Energy loss in the targets and materials of detectors was estimated by a Monte Carlo (MC) simulation based on GEANT4~\cite{geant4}. After the momentum calibration, a missing mass accuracy was estimated to be $\pm1$ ${\rm MeV}/c^{2}$.  \par

 Hyperball-J consisted of 27 coaxial-type Ge detectors having a crystal size of $\rm 70~mm\phi \times 70~mm$. The absolute photopeak efficiency taking the absorption in the target material into account was 3$\%$ for 1 $\rm MeV$ $\gamma$ ray. At least one hit in the Ge detectors was requested to make the trigger. In the off-line analysis, Ge detector events having coincident timing with the $(K^{-},\pi^{-})$ reaction were selected by using a timing gate which depends on $\gamma$-ray energy. Each of the Ge detectors was surrounded by $\rm PbWO_{4}$ (PWO) counters to suppress backgrounds from Compton scattering inside of the Ge crystal and from $\pi^{0}$ decay. We rejected those events in which the Ge detectors have coincident hits with the surrounding PWO counters. \par

 Energy calibration of the Ge detectors was performed by using a natural radioactive $\rm ^{232}Th$ source in the off-beam periods between beam spills and known $\gamma$ rays from nuclei inside the target or the surrounding materials during beam spills. After the energy calibration, we achieved an accuracy of 0.5 keV for the $\gamma$-ray energy range from 0.1 to 2.6 MeV. The energy resolution of the Ge detectors was measured to be 4.5 $\rm keV$ (FWHM) for 1 MeV $\gamma$ ray. Variation of $\gamma$-ray peak shape due to the Doppler broadening effect for the $\rm ^{19}_{\Lambda}F$ in flight was estimated by using a MC simulation based on the SRIM code for deceleration of $\rm ^{19}_{\Lambda}F$ in the $\rm {CF_{4}}$ target~\cite{srim}. We used the results of the MC to distinguish between in-flight and at-rest $\gamma$-ray emission and to assign the $\gamma$ transition. \par

 We selected the reaction angle ($\theta$) range of $2^{\circ} < \theta < 12^{\circ}$, due to large background from the beam decays at angles smaller than $2^{\circ}$ and the small production cross section for $\rm ^{19} _{\Lambda} F$ at angles larger than $12^{\circ}$. In addition, we also acquired data samples with a thin $\rm CF_{2}$ (Teflon) target of $6.6~\rm g/cm^{2}$ thickness to verify the $\rm ^{19}_{\Lambda} F$ production with a better missing mass resolution ($5.9~{\rm MeV}/c^{2}$ FWHM) and without $\gamma$-ray hit bias in the trigger condition. \par


\begin{figure}[t]
\includegraphics[width=0.5\textwidth]{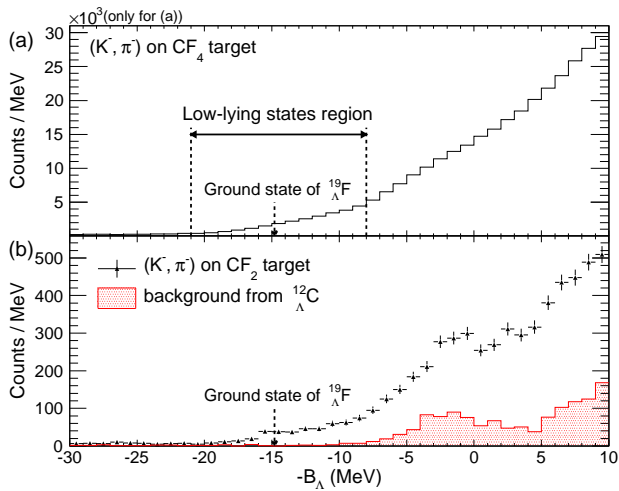}
\caption{Hypernuclear mass spectra of $\rm ^{19}_{\Lambda}F$ with (a) the liquid $\rm CF_{4}$ target (20 $\rm g/cm^{2}$) and (b) the $\rm CF_{2}$ (Teflon) target (6.6 $\rm g/cm^{2}$), plotted in the $\Lambda$ binding energy ($B_{\Lambda}$). In only (a), events are requested to have at least one hit in the Ge detectors. The ground state energy of $\rm ^{19}_{\Lambda}F$ and the ``low-lying states region'' described in the text are marked. In (b), the shaded area indicates the background from $\rm ^{12}_{\Lambda} C$. }

\label{fig:be}
\end{figure}

 Figure~\ref{fig:be} shows the mass spectra for (a) the thick $\rm CF_{4}$ target and (b) the thin $\rm CF_{2}$ target plotted in the $\Lambda$ binding energy ($B_{\Lambda}$) scale. The ground state of $\rm {^{19}_{\Lambda}F}$ is expected to be at $-B_{\Lambda}=-14.8~\rm MeV$~\cite{umeya_theoretical_study}. In Fig.~\ref{fig:be}(b), the spectrum has a significant number of events around the $\rm {^{19}_{\Lambda}F}$ ground state energy and above, indicating the production of low-lying $s_{\Lambda}$ states of $\rm {^{19}_{\Lambda}F}$. In Fig.~\ref{fig:be}(a), we cannot see a clear structure of $\rm ^{19}_{\Lambda} F$ because of 8.7 $\rm MeV$ FWHM resolution. As shown in Fig.~\ref{fig:be}(a), we selected the $B_{\Lambda}$ range of $-21 < -B_{\Lambda} < -8~{\rm MeV}$ (the ``low-lying states region'') in order to observe $\gamma$ rays emitted from the $\rm {^{19}_{\Lambda}F}$ low-lying states as given in Fig.~\ref{fig:scheme}. This region does not cover all the excited states which can populate the low-lying states through $\gamma$ cascades. However, we did not extend the region to avoid background $\gamma$ rays from hyperfragments such as $\rm ^{15}_{\Lambda} N$, $\rm ^{18}_{\Lambda} O$, or $\rm ^{18}_{\Lambda} F$ after $\alpha$, $p$, or $n$ emissions from $p_{\Lambda}$ or highly-excited states of $\rm ^{19}_{\Lambda}F$; the expected energy for the $p_{\Lambda}$ states and the lowest particle emission threshold for hyperfragments are $-B_{\Lambda} \cong -4$ and $-9~{\rm MeV}$, respectively~\cite{umeya_theoretical_study}, and thus all the $p_{\Lambda}$ states of $\rm ^{19}_{\Lambda} F$ decay into hyperfragments. In addition, $\rm ^{12}_{\Lambda}C$ states produced on $\rm ^{12}C$ in the $\rm CF_{4}$ target also contribute to the background at $-B_{\Lambda} \gtrsim -8~{\rm MeV}$ in the $\rm ^{19}_{\Lambda}F$ mass spectrum, as confirmed by the red histogram in Fig.~\ref{fig:be}(b) which was obtained from $\rm CH_{2}$ target data taken with the same setup and analyzed with the ${\rm ^{19}F}(K^{-}, \pi^{-})$ kinematics. \par  

\begin{figure*}[htb]
\includegraphics[width=1.0\textwidth]{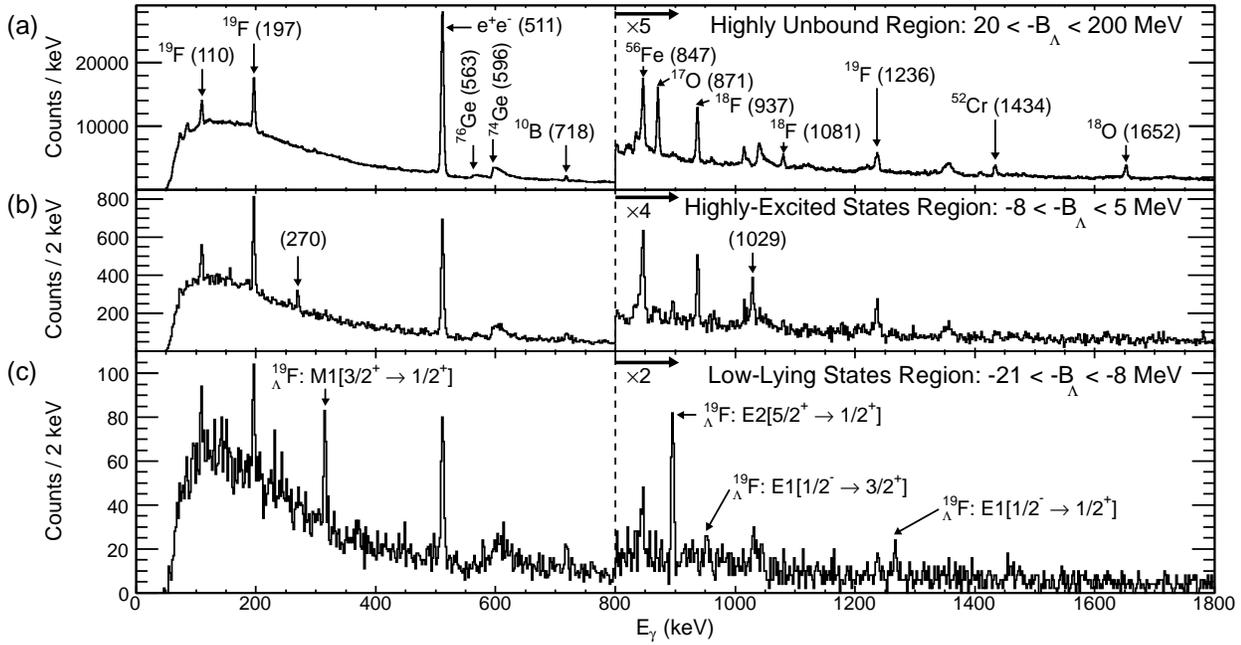}
\caption{$\gamma$-ray energy spectra gated by $\Lambda$ binding energy ranges, (a) unbound region ($20<-B_{\Lambda}<200~\rm MeV$), (b) highly-excited states region ($-8<-B_{\Lambda}<5~\rm MeV$), and (c) low-lying states region ($-21<-B_{\Lambda}<-8~\rm MeV$). Several $\gamma$-ray peaks from ordinary nuclei are marked with their source in (a). Two new $\gamma$-ray peaks at 270 and 1029 $\rm keV$ are shown in (b). In (c), two $\gamma$ rays at 316 and 895 $\rm keV$ are assigned as $M1(3/2^{+} \rightarrow 1/2^{+})$ and $E2(5/2^{+} \rightarrow 1/2^{+})$ transitions, respectively (see text). Other two $\gamma$-ray peaks at 953 and 1266 $\rm keV$, assigned as $E1$ transitions, are magnified in Fig.~\ref{fig:gamma2}.}
\label{fig:gamma1}
\end{figure*}

 Figure~\ref{fig:gamma1} shows the $\gamma$-ray spectra for various $B_{\Lambda}$ regions: (a) the ``highly unbound region'' ($20 <-B_{\Lambda}<200~{\rm{MeV}}$), (b) the ``highly-excited states region'' ($-8 <-B_{\Lambda}<5~{\rm{MeV}}$), and (c) the ``low-lying states region''. Several known $\gamma$ rays from ordinary nuclei such as $\rm^{19}F$, $\rm^{18}F$, etc. are seen in all the spectra as expected for background $\gamma$ rays. The highly-excited states region was selected to identify $\gamma$-ray peaks which were emitted from high excited states including $p_{\Lambda}$ states and hyperfragments. In (b), we observed two unknown $\gamma$-ray peaks with 270 and 1029 $\rm keV$ energies, and their sources are considered as hyperfragments. After gating the low-lying states region, three peaks appeared at 316, 895, and 1266 $\rm keV$ in the $\gamma$-ray spectrum (c), with statistical significances of more than $3\sigma$. \par

 The $\gamma$-ray peak at 316 keV is attributed to the $M1(3/2^{+} \rightarrow 1/2^{+})$ transition between the ground state doublet members, because the yield of the $3/2^{+} \to 1/2^{+}$ $M1$ transition is expected to be more than 10 times larger than the other transitions in the $100 - 500 ~ \rm keV$ energy range~\cite{umeya_theoretical_study}. The energy and width of the $\gamma$-ray peak are determined to be $\rm 315.5 \pm 0.4 (stat) ^{+0.6} _{-0.5} (syst)~keV$  and $\rm 5.0 \pm 0.9 ^{+0.5} _{-0.3}~keV$ (FWHM), respectively. 
\par

 As shown in Fig.~\ref{fig:gamma1}(c), the peak at 895 $\rm keV$ exhibits a narrow width, $\rm 4.3 \pm 0.5 (stat) ^{+0.1} _{-0.2} (syst)~keV$ (FWHM), which is consistent with the energy resolution of the Ge detectors, $4.5 ^{+0.4}_{-0.3} ~\rm keV$ (FWHM). It indicates that the $\gamma$ rays are emitted after the $\rm {^{19}_{\Lambda}F}$ hypernucleus has completely stopped. By comparing it with expected lifetimes and cross sections of $\rm {^{19}_{\Lambda}F}$ states~\cite{umeya_theoretical_study}, this $\gamma$-ray peak is attributed to the $E2(5/2^{+} \rightarrow 1/2^{+})$ or $E1(1/2^{-} \rightarrow 3/2^{+})$ transitions. Here the latter assignment is rejected because $E1(1/2^{-} \rightarrow 1/2^{+})$ is not seen at 1.21 $\rm MeV$ in spite of the expected branching ratio, $BR(1/2^{-} \rightarrow 1/2^{+})/BR(1/2^{-} \rightarrow 3/2^{+})=1.7$. A peak from the $5/2^{+} \to 3/2^{+}$ $\gamma$ transition is not clearly seen at 580 $\rm keV$ in Fig.~\ref{fig:gamma1}(c), but it is consistent with the fact that the $M1(5/2^{+} \rightarrow 3/2^{+})$ transition is largely suppressed in the weak coupling limit. From a fit to the 895 $\rm keV$ peak, the energy of the $E2(5/2^{+} \rightarrow 1/2^{+})$ is derived as $\rm 895.2 \pm 0.3 (stat) \pm 0.5 (syst)~keV$. \par

\begin{figure}[b]
\includegraphics[width=0.5\textwidth]{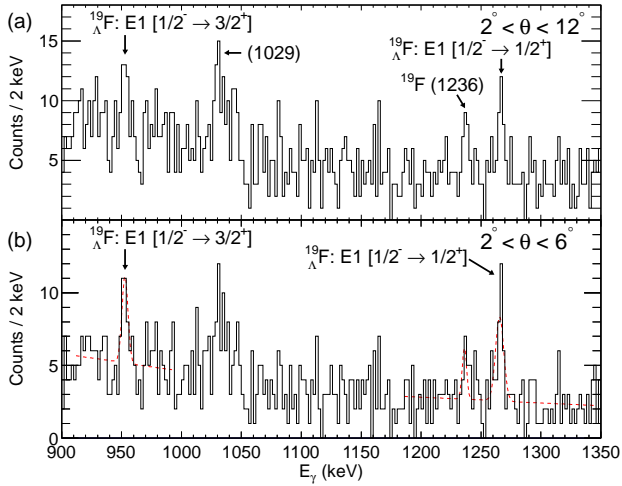}
\caption{$\gamma$-ray spectra for the low-lying states region, $-21<-B_{\Lambda}<-8~\rm MeV$. (a) shows the enlargement of Fig.~\ref{fig:gamma1}(c) from 900 to 1350 $\rm keV$. In (b), the forward reaction angles of $2^{\circ} < \theta < 6^{\circ}$ are selected. Two $\gamma$-ray peaks are observed at 953 and 1266 $\rm keV$, which are assigned as $E1(1/2^{-} \rightarrow 3/2^{+})$ and $E1(1/2^{-} \rightarrow 1/2^{+})$, respectively. The dotted lines in (b) show fitting results.}
\label{fig:gamma2}
\end{figure}

 We considered the $\gamma$ transition of the 1266 $\rm keV$ peak in Fig.~\ref{fig:gamma1}(c) as the $E1(1/2^{-} \rightarrow 1/2^{+})$ due to the expected large cross section of the $1/2^{-}$ state~\cite{umeya_theoretical_study}. To confirm this assignment, the events with forward reaction angles of $2^{\circ} - 6^{\circ}$ were selected, since the cross section of the $1/2^{-}$ state is expected to be the largest at $4^{\circ}$~\cite{umeya_theoretical_study}. At the forward reaction angles, as shown in Fig.~\ref{fig:gamma2}(b), the peak is more evident and another peak at 953 $\rm keV$ is also seen with a statistical significance of 3.2$\sigma$. From a fit to the $\gamma$ ray spectrum, the energies of the two peaks are determined to be $\rm 952.8 \pm 1.2 (stat) ^{+0.5} _{-0.6} (syst)~keV $ and $\rm 1265.6 \pm 1.2 (stat) ^{+0.7} _{-0.5} (syst)~keV$. The energy difference between the two peaks, $\rm 312.7 \pm 1.7 {\rm (stat)}~keV$, is consistent with the energy spacing between the ground state doublet, which is found to be $\rm 315.5 \pm 0.4 {\rm (stat)} ~keV$ as described above. It indicates that the 953 and 1266 $\rm keV$ $\gamma$ rays are emitted from the same initial state decaying to the $3/2^{+}$ and $1/2^{+}$ states, respectively. Therefore, we assigned the $\gamma$ transitions as $E1(1/2^{-} \rightarrow 3/2^{+})$ and  $E1(1/2^{-} \rightarrow 1/2^{+})$. The measured yields and the widths of the $\gamma$ rays support this assignment. By using the calculated transition probabilities of $B(E1;1/2^{-} \rightarrow 3/2^{+},1/2^{+})$~\cite{umeya_theoretical_study}, the yield of the 953 $\rm keV$ $\gamma$ ray was estimated to be $18 \pm 6$ events from the measured yield of the 1266 $\rm keV$ $\gamma$ ray, $\rm 25 \pm 8 (stat) ^{+2} _{-5}(syst) $ events, and a relative Hyperball-J efficiency for the different $\gamma$-ray energies. The estimated yield is consistent with the measured yield, $\rm 19 \pm 8 (stat) ^{+1} _{-2}(syst)$ events. Additionally, the widths of the 953 and 1266 keV peaks, $\rm 5.9 \pm 2.1 (stat) ^{+0.2} _{-0.1} (syst)~keV$ and $\rm 8.1 \pm 3.0 (stat) ^{+0.5} _{-0.6} (syst)~keV$ (FWHM), respectively, are mainly attributed to the detector resolution [$4.5^{+0.4}_{-0.3}$ keV and $4.7^{+0.4}_{-0.3}$ keV (FWHM), respectively]. Because of the very slow core transition of $\rm ^{18}F(0^{-} \to 1^{+})$, these $E1$ transitions of $\rm ^{19}_{\Lambda}F$ are almost unbroadened by Doppler shift, while the $M1(1/2^{+}(T=1) \rightarrow 3/2^{+}, 1/2^{+})$ transitions in Fig.~\ref{fig:scheme} are Doppler broadened and expected to have widths of $\rm 10.8 ^{+0.5} _{-0.3} ~ keV$ and $\rm 14.0 ^{+0.7} _{-0.4} ~ keV$ (FWHM), respectively, according to the MC simulation.
\par

 The measured $(3/2^{+}, 1/2^{+})$ doublet spacing of 316 keV is in good agreement with the two independent shell-model calculations so far. It is noted that both calculations reproduce the spacing energies of spin-spin doublets for $s$- and $p$-shell hypernuclei. Millener predicted a spacing energy to be 305 keV from the phenomenological spin-dependent $\Lambda N$ interaction strengths determined from the $p$-shell hypernuclear data~\cite{millener_theoretical_study}. Since the $\Lambda \Sigma$ coupling effect is not included in this calculation, the agreement suggests that the $\Lambda \Sigma$ coupling effect to the hypernuclear spin-spin doublet spacing, which is significantly large in the $s$-shell hypernuclei~\cite{Akaishi_theoretical_study} and smaller in the $p$-shell hypernuclei~\cite{millener_theoretical_study, Wirth_three_body_study}, is negligibly small in the $sd$-shell hypernuclei. \par

 On the other hand, the shell-model calculation by Umeya and Motoba with the effective $\Lambda N$ interaction obtained from the Nijmegen SC97e and SC97f interactions via G-matrix method predicts the spacing of 245 keV~\cite{umeya_theoretical_study_private} and 419 keV~\cite{umeya_theoretical_study}, respectively. The situation is quite similar to the case of the A=4 and 7 hypernuclei. For the ground-state doublet spacing of $\rm {^{7}_{\Lambda}Li}$ ($1/2^{+},3/2^{+}$; 0.692 MeV)~\cite{Tamura_hyperball_7_lambda_li}, the Nijmegen G-matrix interaction gives 0.348 MeV (SC97e) and 0.742 MeV (SC97f)~\cite{umeya_theoretical_study_private}, and for the spacings of $\rm ^{4}_{\Lambda}H$ and $\rm ^{4}_{\Lambda}He$ ($0^{+},1^{+}$; 1.25 MeV in average for $\rm ^{4}_{\Lambda}H$ and $\rm ^{4}_{\Lambda}He$)~\cite{Yamamoto_hyperball_4_lambda_he}, it gives 0.89 MeV (SC97e) and 1.48 MeV (SC97f)~\cite{Akaishi_theoretical_study}. In addition, these agreements indicate that the weak coupling assumption between a $0s$-state $\Lambda$ and the core nucleus, one of the most basic concepts in $\Lambda$-hypernuclear structure, is still valid in $sd$-shell hypernuclei.
\par

 In summary, we observed $\gamma$ rays from an $sd$-shell $\Lambda$ hypernucleus, $\rm ^{19} _{\Lambda}F$, for the first time. The energy spacing between the $3/2^{+}$ and $1/2^{+}$ states is determined to be $\rm 315.5 \pm 0.4 (stat) ^{+0.6} _{-0.5} (syst)~keV$. We also determined the excitation energies of the $5/2^{+}$ and $1/2^{-}$ states to be $\rm 895.2 \pm 0.3 (stat) \pm 0.5 (syst)~keV$ and $\rm 1265.6 \pm 1.2 (stat) ^{+0.7}_{-0.5} (syst)~keV$, respectively. The ($3/2^{+}, 1/2^{+}$) energy spacing is well reproduced by the $\Lambda N$ spin-dependent interactions which reproduced the $s$- and $p$-shell hypernuclear data. It also suggests that the $\Lambda \Sigma$ coupling effect is diminished in heavier hypernuclei. Our result shows that the present theoretical frameworks work quite successfully in describing structure of not only light $s$- and $p$-shell hypernuclei but also a heavier one beyond $p$-shell hypernuclei. Such precise spectroscopic studies of light to heavy $\Lambda$ hypernuclei would also provide unique means to investigate nuclear density dependence of the baryon-baryon interactions in nuclear matter. 
\par

\begin{acknowledgments}
 We thank the J-PARC accelerator and facility staff and all the project members who participated in the development of SKS and Hyperball-J. We also thank A. Umeya, T. Motoba and D. J. Millener for their theoretical calculations of $\rm ^{19} _{\Lambda} F$. This study is supported by Grants-in-Aid Nos. 17070001, 21684011, 23244043, 24105003, and 15H02079 for Scientific Research from the Ministry of Education of Japan. The research is also supported by WCU Grant No. R32-10155, NRF Grant No. 2010-0004752, and Center for Korean J-PARC Users Grant No. K2100200173811B130002410 in Korea. \par
\end{acknowledgments}


\end{document}